\documentclass[a4paper,11pt]{article}
\pdfoutput=1 

\usepackage{paperpub} 

\usepackage[T1]{fontenc} 

\usepackage{subfigure}

\title{\boldmath Holographic calculations of R\'{e}nyi entropy from rotating topological black holes }

\author{Bin He}

\affiliation{Department of Physics and Center for Field Theory and Particle Physics,\\Fudan University,\\Shanghai 200433, China}

\emailAdd{22210190010@m.fudan.edu.cn}

\abstract{We construct a new class of entanglement measures named "rotating R\'{e}nyi entropy" by the holographic calculation of a rotating topological black hole belonging to the Petrov type-D class, we compute this kind of entropy for a spherical entangling surface of a vacuum state in the dual CFT. We find that the latter one could be conformally transformed to a thermal state on the boundary geometry of the rotating topological black hole with certain temperature. Then according to the angular velocity of the bulk black hole, we introduce the chemical potential in the boundary thermodynamics, which appears in the definition of the generalized R\'{e}nyi entropy, and we presented the behaviors of this generalized R\'{e}nyi entropy with the chemical potential and replica parameters.}

\begin{document} 
\maketitle
\flushbottom

\section{Introduction}
\label{sec:intro}

"Topological black holes" is a kind of black holes whose event horizon have nontrivial topological structures, they exist in Anti de-Sitter (AdS) spacetime. In Einstein gravity, they can be expressed as a vacuum solution of the Einstein’s field equations with negative cosmological constant. The event horizon with arbitrary topological structure could be constructed by modding out discrete isometric groups in four-dimensional spacetime~\cite{1,2,3,4,5,6},which is the origin of the name "topological black hole". In this paper, we introduce a rotating topological black hole belonging to Petrov type-D class, its event horizon is a rotating hyperbolic surface. This kind of black hole can be obtained by the analytical continuation of Kerr-de Sitter metric~\cite{7}. According to the AdS/CFT correspondence, we can use this rotating topological black hole to find some physical quantities in the dual conformal field theory (CFT) holographically. \\
We recall the common definition of R\'{e}nyi entropy. If we consider the quantum system with two components A and B, where A, B are spatial regions separated by a co-dimension two entangling surface. We trace over the degrees of freedom in region B and define the reduced density matrix
\begin{equation}
\begin{split}
\rho_A =\rm{Tr}_B\, \rho_{tot}.
\end{split}
\end{equation}
Then we can define the R\'{e}nyi entropy according to this density matrix as
\begin{equation}
\label{eq:1.2}
\begin{split}
S_n=\dfrac{1}{1-n}\,\rm{log\, Tr}\, \rho_A^n.
\end{split}
\end{equation}
According to the definition, the limit $n \to1$ gives the entanglement entropy
\begin{equation}
\begin{split}
S_{EE}=\lim_{n \to1}S_n=-\rm{Tr}\,\rho_A\,\rm{log}\,\rho_A.
\end{split}
\end{equation}
Actually, we can also define these entropies at finite temperature $T=\beta^{-1}$, to do this we can simply replace the reduced density matrix as the thermal density matrix $\rho_{therm}$, where the definition of thermal density matrix is
\begin{equation}
\begin{split}
\rho_{therm}=\frac{\mathrm{e}^{-\beta H}}{\mathrm{Tr}\left({\mathrm{e}^{-\beta H}}\right)}.
\end{split}
\end{equation}
where $H$ is the total Hamiltonian of the system. By definition, if the subsystem equals to the hole system, the thermal entropy is equal to the entanglement entropy at temperature $T$, $i.e.$, $S_A(T)=S_{therm}$. In quantum field theory, R\'{e}nyi entropies can be constructed through the replica trick. According to~\cite{8}, $\rm{Tr}\,\rho_A^n$ can be calculated by the Euclidean path integral on an n-sheeted geometry 
\begin{equation}
\begin{split}
\mathrm{Tr}\,\rho_A^n=Z_1^{-n}\int_{(t_E,x)\in R_n}{D\phi\, e^{-S(\phi)}}=\frac{Z_n}{(Z_1)^n}.
\end{split}
\end{equation}
then \eqref{eq:1.2} could be written as
\begin{equation}
\begin{split}
S_n=\frac{1}{n-1}\,(n\,\mathrm{log}Z_1-\mathrm{log}Z_n).
\end{split}
\end{equation}
In the main discussion of this paper, we will consider the CFT with a conserved angular momentum, which is introduced in the following “grand canonical” generalization of the R\'{e}nyi entropy. Firstly, we generalize the reduced density matrix as\\
\begin{equation}
\begin{split}
\rho_A\to \rho_A\,\frac{\mathrm{e}^{ \mu J_A}}{n(\mu)}.
\end{split}
\end{equation}
Here $\mu$ is the chemical potential conjugate to the angular momentum, and the definition of the normalization factor $n(\mu)$ is
\begin{equation}
\begin{split}
n(\mu)=\mathrm{Tr}\,\left(\mathrm{e}^{\mu J_A}\right).
\end{split}
\end{equation}
After this, we can generalize eq.\eqref{eq:1.2} to the form
\begin{equation}
\begin{split}
S_n(\mu)=\frac{1}{1-n}\,\mathrm{log\,Tr}\left [\rho_A\,\frac{\mathrm{e}^{ \mu J_A}}{n(\mu)}\right]^n.
\end{split}
\end{equation}
the path integral calculations of R\'{e}nyi entropy introduced above can easily extended to compute our new rotating R\'{e}nyi entropy. However, in the holographic framework, we can make it more easily. In this standard framework, the black hole in AdS could be equally described as the thermal state in dual CFT, the latter is defined on the background geometry associated with the geometric conformal boundary of AdS spacetime. Next, if we consider the reduced density matrix of the CFT vacuum state on the flat spacetime across a spherical entangling surface, then it could be found that the latter can be related to a thermal density matrix of the CFT on the boundary of the rotating topological black hole with a rotating chemical potential.\\
The layout of this paper is as follows: in Section \ref{sec:2} we introduce the calculations of the rotating topological black hole, we give out the Bekenstein-Hawking entropy and Hawking temperature of the black hole, and also the angular velocity of the AdS bulk. In Section \ref{sec:3}, we consider the relations between the CFT vacuum state on flat geometry and the CFT thermal state on the geometry of the rotating black hole, and according to standard AdS/CFT we use the black hole to calculate the entropy and temperature of the CFT on the latter geometry, then the entropies of the former also calculated. We make the conclusions and discussions in Section \ref{sec:4}.

\section{The calculation of AdS}
\label{sec:2}
\subsection{Rotating topological black holes}
We firstly introduce the metric of the rotating topological black hole, in Boyer-Lindquist-type coordinates its metric reads
\begin{equation}
\label{eq:2.1}
\begin{split}
ds^2=&\frac{1}{\rho^2}\left( \Delta_\theta a^2\mathrm{sinh}^2\,\theta-\Delta_r\right)dt^2+\frac{\rho^2}{\Delta_r}dr^2+\frac{\rho^2}{\Delta_\theta}d\theta^2\\
&+\frac{\mathrm{sinh}^2\,\theta}{\rho^2\Xi^2}\left[\Delta_\theta \left(r^2+a^2\right)^2-\Delta_ra^2\mathrm{sinh}^2\,\theta\right]d\phi^2-2\frac{a\,\mathrm{sinh}^2\,\theta}{\rho^2\Xi}\left[\Delta_\theta \left(r^2+a^2\right)+\Delta_r\right]dtd\phi.
\end{split}
\end{equation}
where
\begin{equation}
\begin{split}
\rho^2=&r^2+a^2\,\mathrm{cosh}^2\,\theta,\\
\Delta_r=&\left(r^2+a^2\right)\left(-1-\frac{\Lambda r^2}{3}\right)-2\eta r,\\
\Delta_\theta=&1-\frac{\Lambda a^2}{3}\,\mathrm{cosh}^2\,\theta,\\
\Xi=&1-\frac{\Lambda a^2}{3}.
\end{split}
\end{equation}
in them $\eta$ is the mass parameter, $a$ is the rotational parameter, and $\Lambda=-{l^2}/{3}$ is the cosmological constant. It can be seen that due to the spacetime cross term in the metric \eqref{eq:2.1}, the spacetime it describes is a non-static one. According to ~\cite{7}, metric \eqref{eq:2.1} could be obtained by the analytical continuation of Kerr-de Sitter metric. In Boyer- Lindquist-type coordinates, its form reads 
\begin{equation}
\label{eq:2.3}
\begin{split}
ds^2=&\frac{1}{\rho^2}\left( \Delta_\theta a^2\mathrm{sin}^2\,\theta-\Delta_r\right)dt^2+\frac{\rho^2}{\Delta_r}dr^2+\frac{\rho^2}{\Delta_\theta}d\theta^2\\
&+\frac{\mathrm{sin}^2\,\theta}{\rho^2\Xi^2}\left[\Delta_\theta \left(r^2+a^2\right)^2-\Delta_ra^2\mathrm{sin}^2\,\theta\right]d\phi^2-2\frac{a\,\mathrm{sin}^2\,\theta}{\rho^2\Xi}\left[\Delta_\theta \left(r^2+a^2\right)+\Delta_r\right]dtd\phi.
\end{split}
\end{equation}
where now
\begin{equation}
\begin{split}
\rho^2=&r^2+a^2\,\mathrm{cos}^2\,\theta,\\
\Delta_r=&\left(r^2+a^2\right)\left(1-\frac{\Lambda r^2}{3}\right)-2\eta r,\\
\Delta_\theta=&1+\frac{\Lambda a^2}{3}\,\mathrm{cos}^2\,\theta,\\
\Xi=&1+\frac{\Lambda a^2}{3}.
\end{split}
\end{equation}
Now we note that \eqref{eq:2.1} can be obtained from \eqref{eq:2.3} by the analytical continuation
\begin{equation}
\label{eq:2.5}
\begin{split}
&t\to it,\quad r\to ir,\quad \theta \to i\theta,\quad \phi \to \phi,\\
&\eta \to -i\eta,\quad a\to ia.
\end{split}
\end{equation}
where the sign of $\Lambda$ is also changed. \\
It is easily to observes that the metric \eqref{eq:2.1} describes a spacetime in the limit $a=0$ reduced to the static topological black hole. In order to verify that formula \eqref{eq:2.1} is indeed the solution of Einstein’s equations, we consider the solution of the most general Petrov type-D Einstein Maxwell equations with negative cosmological constant. After rescaling, we get its metric as
\begin{equation}
\label{eq:2.6}
\begin{split}
ds^2=&\frac{p^2+q^2}{P}dp^2+\frac{P}{p^2+q^2}\left(d\tau+q^2d\sigma\right)^2+\frac{p^2+q^2}{L}dq^2\\
&-\frac{L}{p^2+q^2}\left(d\tau-p^2d\sigma\right)^2.
\end{split}
\end{equation}
where
\begin{equation}
\begin{split}
P=&\gamma+2np-\varepsilon p^2-\frac{\Lambda}{3}p^4,\\
L=&\gamma-2\eta q+\varepsilon q^2-\frac{\Lambda}{3}q^4.
\end{split}
\end{equation}
Replace the following parameters
\begin{equation}
\begin{split}
q=&r,\quad p=a\mathrm{cosh}\,\theta,\quad \tau=t-a\phi,\quad \sigma=-\frac{\phi}{a},\\
\varepsilon=&-1-\frac{\Lambda a^2}{3}, \quad \gamma=-a^2, \quad n=0.
\end{split}
\end{equation}
In this way, the metric \eqref{eq:2.1} is obtained. This shows that the result of the metric \eqref{eq:2.1} is a limit case of the more general solution \eqref{eq:2.6} of Einstein’s equations. This means that formula \eqref{eq:2.1} is indeed the solution of Einstein’s equations with cosmological constant, and it also means that the analytical continuation \eqref{eq:2.5} of Kerr-de Sitter metric is also a possible solution.
\subsection{The calculation of entropy and temperature}
Next, we calculate the Bekenstein-Hawking entropy of the rotating topological black hole described in metric \eqref{eq:2.1}, which means that we need to calculate the area of the event horizon of the black hole. According to the metric of AdS spacetime, fixing the radial coordinate $r$ will induce a time-like 3-surface, and then fixing the coordinate time $t$, we will foliate the surface into a series of space-like 2-surfaces. The metric induced on these surfaces is:
\begin{equation}
\label{eq:2.9}
\begin{split}
d\sigma^2=\frac{\rho^2}{\Delta_\theta}d\theta^2+ \frac{\Sigma^2\,\mathrm{sinh}^2\,\theta}{\rho^2\Xi^2}d\phi^2 .
\end{split}
\end{equation}
where we introduced
\begin{equation}
\begin{split}
\Sigma^2 \equiv \Delta_\theta \left(r^2+a^2\right)^2-\Delta_ra^2\mathrm{sinh}^2\,\theta.
\end{split}
\end{equation}
We assume that the equation describes the event horizon is $f(x^\mu)=0$, according to the metric \eqref{eq:2.1}, considering its symmetry in the direction of coordinates $t$ and $\phi$, so we have $f(x^\mu)=f(r,\theta)$. The event horizon of a black hole is defined as a surface whose normal unit vector have zero length, $i.e.$, a zero-hypersurface in the spacetime. By definition we can write
\begin{equation}
\begin{split}
n^\mu n_\mu=g^{\mu \nu} \partial_\mu f\partial_\nu f=0.
\end{split}
\end{equation}
By separating the variables of this equation, we get the equation of the horizon
\begin{equation}
\begin{split}
\frac{\Delta_r}{R^2}\left(\frac{dR}{dr}\right)^2=0.
\end{split}
\end{equation}
which means that 
\begin{equation}
\label{eq:2.13}
\begin{split}
\left(r^2+a^2\right)\left(\frac{r^2}{l^2}-1\right)-2\eta r=0.
\end{split}
\end{equation}
The solution of $r_H$ is discussed in detail in ~\cite{7}, in the all cases, the outermost $r_H$ represents the event horizon. Under this condition, the Gaussian curvature of the surface described in equation \eqref{eq:2.9} is
\begin{equation}
\begin{split}
K=-\frac{1}{\rho_H^6}\left\lbrace\left(\rho_H^2-4a^2\,\mathrm{cosh}^2\,\theta \right)\left[\left(r_H^2+a^2\right)\Delta_\theta +\frac{\rho_H^2a^2}{l^2}\,\mathrm{sinh}^2\,\theta \right]+\frac{4\rho_H^2a^2}{l^2}\,\mathrm{cosh}^2\,\theta\right\rbrace.
\end{split}
\end{equation}
The indicator $H$ indicates the value at the event horizon. Unlike the case of non-rotation, K is no longer constant because the event horizon is deformed due to the rotation. Then we calculate the area of the event horizon, to do this we combine \eqref{eq:2.9} and \eqref{eq:2.13}
\begin{equation}
\begin{split}
A_H=\int d\phi d\theta \left.\sqrt{h}\right|_{\Delta_r=0}=V_H l^2\, \frac{r_H^2+a^2}{l^2+a^2}.
\end{split}
\end{equation}
where we introduced $V_H\equiv\int_0^{\theta_m}d\theta\,\mathrm{sinh}\,\theta \int_0^{2\pi}d\phi$. According to the Bekenstein-Hawking formula, the horizon entropy is
\begin{equation}
\label{eq:2.16}
\begin{split}
S_{BH}=\frac{V_H l^2}{4G_N}\, \frac{r_H^2+a^2}{l^2+a^2}.
\end{split}
\end{equation}
Next, we consider the Hawking temperature of the black hole, which can be determined by the following formula
\begin{equation}
\begin{split}
T_H=\frac{\kappa}{2\pi}.
\end{split}
\end{equation}
where $\kappa$ is the surface gravity of the event horizon. By definition, the surface gravity can be calculated by the following method 
\begin{equation}
\begin{split}
\kappa=\lim_{r\to r_H}\left(a^{(4)}\sqrt{-g_{00}}\right).
\end{split}
\end{equation}
where $a^{(4)}$ is the proper acceleration of the stationary particle near the event horizon of the black hole. In order to calculate this quantity, we introduce the so-called dragging frame, in which the metric \eqref{eq:2.1} is
\begin{equation}
\label{eq:2.19}
\begin{split}
ds^2=-\frac{\rho^2\Delta_\theta \Delta_r}{\Sigma^2}dt^2+\frac{\rho^2}{\Delta_r}dr^2+\frac{\rho^2}{\Delta_\theta}d\theta^2.
\end{split}
\end{equation}
this coordinate system will rotate with the black hole at angular velocity
\begin{equation}
\begin{split}
\omega=\frac{a\Xi\left[\left(r^2+a^2\right)\Delta_\theta+\Delta_r\right]}{\Sigma^2}.
\end{split}
\end{equation}
In the spacetime described in \eqref{eq:2.19}, we can calculate that the surface gravity is
\begin{equation}
\begin{split}
\kappa=\frac{1}{2\left(r_H^2+a^2\right)l^2}\left[3r_H^3+\left(a^2-l^2\right)r_H+\frac{a^2l^2}{r_H}\right].
\end{split}
\end{equation}
Then the temperature
\begin{equation}
\label{eq:2.22}
\begin{split}
T_H=\frac{\kappa}{2\pi}=\frac{1}{4\pi \left(r_H^2+a^2\right)l^2}\left[3r_H^3+\left(a^2-l^2\right)r_H+\frac{a^2l^2}{r_H}\right].
\end{split}
\end{equation}
It is worth noting that although the metric \eqref{eq:2.1} does not have all rotational symmetry, the temperature \eqref{eq:2.22} is still a constant on the event horizon. We can find that if the mass parameter $\eta=0$, then we can find out $T=1/{2\pi l}$, which is very meaningful for the further consideration between the rotation and stationary.

\section{The holographic calculation of CFT}
\label{sec:3}
\subsection{Relations between the vacuum and finite temperature}
In this section, we will focus on the computations of rotating R\'{e}nyi entropy in 3-dimensional conformal field theories. To do this we consider a CFT vacuum state in flat space, where the entangling surface is a co-dimension two sphere of radius $R$ in a constant time slice. In this case, according to the argument of ref.~\cite{9}, the entanglement entropy across this entanglement surface equals the thermal entropy of the CFT on a hyperbolic cylinder, then the temperature and curvature are determined by the radius of the original entangling surface. After some work, we find that this approach also can be used to construct our rotating R\'{e}nyi entropy. Firstly, we give a quick review of this approach, we start with the flat space metric in polar coordinates:
\begin{equation}
\label{eq:3.1}
\begin{split}
ds^2=-dt^2+dr^2+r^2d\phi^2.
\end{split}
\end{equation}
Our entangling surface $\Sigma$ is again the sphere $r=R$ on the constant time slice. Then we consider the following coordinate transformation
\begin{equation}
\begin{split}
t=&R\frac{\mathrm{sinh}\left(\tau/R\right)}{\mathrm{cosh}\,\theta+\mathrm{cosh}\left(\tau/R\right)},\\
r=&R\frac{\mathrm{sinh}\,\theta}{\mathrm{cosh}\,\theta+\mathrm{cosh}\left(\tau/R\right)}.
\end{split}
\end{equation}
One can easily verify the above metric \eqref{eq:3.1} becomes
\begin{equation}
\label{eq:3.3}
\begin{split}
ds^2=\Omega^2\left[-d\tau^2+R^2\left(d\theta^2+\mathrm{sinh}^2\,\theta d\phi^2\right)\right].
\end{split}
\end{equation}
where the conformal pre-factor
\begin{equation}
\begin{split}
\Omega=\left[\mathrm{cosh}\,\theta+\mathrm{cosh}\left(\tau/R\right)\right]^{-1}.
\end{split}
\end{equation}
We find that, after remove the pre-factor $\Omega$, the conformal transformation maps the metric on flat space to one on hyperbolic space which have curvature radius $R$. The key point is that, under this conformal transformation, the density matrix describing the CFT vacuum state at the interior of the spherical entangling surface is transformed to a thermal density matrix on the hyperbolic geometry with the temperature
\begin{equation}
\begin{split}
T_0=\frac{1}{2\pi R}.
\end{split}
\end{equation}
This means that, on the hyperbolic geometry we have a CFT thermal state with the density matrix
\begin{equation}
\label{eq:3.6}
\begin{split}
\rho_{therm}=\frac{\mathrm{e}^{-H/T_0}}{Z(T_0)}.
\end{split}
\end{equation}
In the standard AdS/CFT framework, we can make a holographic calculation of CFT, which means that evaluate the thermal entropy on the transformed hyperbolic background as the horizon entropy of a topological black hole in the bulk with a hyperbolic event horizon. For our purpose, it is of great significance to follow this approach, and the only difference is we have a rotating hyperbolic background as the boundary of the bulk.

\subsection{Relations between the static and rotating background}
To begin with, we consider the standard AdS metric
\begin{equation}
\label{eq:3.7}
\begin{split}
ds^2=-\left(\frac{y^2}{l^2}-1\right)dT^2+\left(\frac{y^2}{l^2}-1\right)^{-1}dy^2+y^2\left(d\Theta^2+\mathrm{sinh}^2\,\Theta d\Phi^2\right).
\end{split}
\end{equation}
We note that if the following coordinate transformation is introduced
\begin{equation}
\label{eq:3.8}
\begin{split}
&T=t,\quad \Phi=\phi-\frac{a}{l^2}t,\\
&y\mathrm{cosh}\,\Theta=r\mathrm{cosh}\, \theta,\\
&y^2=\frac{1}{\Xi}\left(r^2\Delta_\theta -a^2\,\mathrm{sinh}^2\,\theta\right).
\end{split}
\end{equation}
Then the standard AdS metric \eqref{eq:3.7} is transformed to the metric of the rotating topological black hole with $\eta=0$. This means that the metric \eqref{eq:2.1} with $\eta=0$ describes the standard AdS spacetime observed by a rotating observer. But in the rotating black hole metric proposed in Section \ref{sec:2}, the induced metric on its asymptotic boundary is the same whether the black hole has mass or not, this can be seen in the asymptotic behavior of the metric:
\begin{equation}
\begin{split}
\rho^2\to r^2,\quad \Delta_r\to \frac{r^4}{l^2},\quad when\quad r\to \infty.
\end{split}
\end{equation}
and the boundary metric
\begin{equation}
\label{eq:3.10}
\begin{split}
ds_{boundary}^2=-\frac{1}{\Xi^2}dt^2+\frac{l^2}{\Delta_\theta}d\theta^2+\frac{l^2}{\Xi}\,\mathrm{sinh}^2\,\theta \,d\phi^2-\frac{2a\,\mathrm{sinh}^2\,\theta}{\Xi^2}dtd\phi.
\end{split}
\end{equation}
But we know that the boundary of metric \eqref{eq:3.7} is 
\begin{equation}
\label{eq:3.11}
\begin{split}
ds_{boundary}^2=-dT^2+l^2\left(d\Theta^2+\mathrm{sinh}^2\,\Theta d\Phi^2\right).
\end{split}
\end{equation}
Which is a standard hyperbolic background. So this tells us the coordinate transformation \eqref{eq:3.8} connects the boundary geometry \eqref{eq:3.10} and \eqref{eq:3.11}, but the latter is exactly the geometry on \eqref{eq:3.3} with $R=l$. In order to verify this approach, we reconsider the temperature of the rotating topological black hole \eqref{eq:2.22}, we have discussed that in the case $\eta=0$, we can find that $T=1/{2\pi l}$. All the above discussions tell us, the conclusion of ref.~\cite{9} is also satisfied when the latter thermal CFT is on a rotating hyperbolic background.
\subsection{Generalized rotating R\'{e}nyi entropies}
Then let us consider the Renyi entropies with a rotating chemical potential. Firstly, we generalize the reduced density matrix as
\begin{equation}
\label{eq:3.12}
\begin{split}
\rho_A\to \rho_A\,\frac{\mathrm{e}^{ \mu J_A}}{n(\mu)}.
\end{split}
\end{equation}
Here $\mu$ is the chemical potential conjugate to the angular momentum. Then the generalized R\'{e}nyi entropy could be written as 
\begin{equation}
\label{eq:3.13}
\begin{split}
S_n(\mu)=\frac{1}{1-n}\,\mathrm{log\,Tr}\left [\rho_A\,\frac{\mathrm{e}^{ \mu J_A}}{n(\mu)}\right]^n.
\end{split}
\end{equation}
According to the replica trick under a grand canonical version, we have
\begin{equation}
\begin{split}
\mathrm{Tr}\left [\rho_A\,\frac{\mathrm{e}^{ \mu J_A}}{n(\mu)}\right]^n=\frac{Z_n(\mu)}{\left[Z_1(\mu)^n\right]}.
\end{split}
\end{equation}
Then the R\'{e}nyi entropy \eqref{eq:3.13} can be written as 
\begin{equation}
\begin{split}
S_n(\mu)=\frac{1}{n-1}\,\left[n\,\mathrm{log}Z_1(\mu)-\mathrm{log}Z_n(\mu)\right].
\end{split}
\end{equation}
We can compute rotating R\'{e}nyi entropies \eqref{eq:3.13} by generalizing the approach discussed above. Then the correspond formula of \eqref{eq:3.6} is
\begin{equation}
\label{eq:3.16}
\begin{split}
\rho_{therm}=\frac{\mathrm{e}^{-H/T_0+\mu J}}{Z(T_0,\mu)}.
\end{split}
\end{equation}
Where we have identified the appropriate angular momentum $J$, and we defined 
\begin{equation}
\begin{split}
T_0=\frac{1}{2\pi l}.
\end{split}
\end{equation}
Then the generalized reduced density matrix \eqref{eq:3.12} and the thermal density matrix on the boundary of rotating topological black hole \eqref{eq:3.16} have the relation 
\begin{equation}
\begin{split}
\rho_A\,\frac{\mathrm{e}^{ \mu J_A}}{n(\mu)}=U^{-1}\rho_{therm}U=U^{-1}\frac{\mathrm{e}^{-H/T_0+\mu J}}{Z(T_0,\mu)}U.
\end{split}
\end{equation}
Where $U$ is the unitary operator corresponding to the conformal transformation between the flat geometric background and the background of AdS asymptotic boundary. Then we have
\begin{equation}
\begin{split}
\left [\rho_A\,\frac{\mathrm{e}^{ \mu J_A}}{n(\mu)}\right]^n=U^{-1}\frac{\mathrm{e}^{-H/(T_0/n)+n\mu J}}{Z(T_0,\mu)^n}U.
\end{split}
\end{equation}
Take trace of the both side
\begin{equation}
\begin{split}
\mathrm{Tr}\left [\rho_A\,\frac{\mathrm{e}^{ \mu J_A}}{n(\mu)}\right]^n=&\mathrm{Tr}\left [U^{-1}\frac{\mathrm{e}^{-H/(T_0/n)+n\mu J}}{Z(T_0,\mu)^n}U\right]\\
=&\frac{Z(T_0/n,\mu)}{Z(T_0,\mu)^n}.
\end{split}
\end{equation}
Then we can write the generalized R\'{e}nyi entropy as
\begin{equation}
\begin{split}
S_n(\mu)=\frac{1}{1-n}\,\mathrm{log}\,\frac{Z(T_0/n,\mu)}{Z(T_0,\mu)^n}.
\end{split}
\end{equation}
In the grand canonical ensemble, according to the relationship between thermodynamic entropy and free energy, we can write
\begin{equation}
\begin{split}
S_{therm}(T,\mu)=&-\frac{ \partial F(T,\mu)}{\partial T}\bigg|_\mu\\
=&\frac{\partial}{\partial T}[T\,\mathrm{log}\,Z(T,\mu)]\bigg|_\mu.
\end{split}
\end{equation}
So the relationship between the generalized R\'{e}nyi entropy and the thermodynamic entropy is
\begin{equation}
\label{eq:3.23}
\begin{split}
S_n(\mu)=\frac{n}{n-1}\,\frac{1}{T_0}\int_{T_0/n}^{T_0}S_{therm}(T,\mu)dT.
\end{split}
\end{equation}

So far, we note that the above discussion does not involve AdS/CFT correspondence. As we take this into the consideration, we can calculate the thermal entropy and temperature as the black hole entropy and temperature. According to ref.~\cite{10}, the chemical potential in the boundary thermodynamics can be constructed through the angular velocity of the bulk. We recall that the spacetime describes the black hole in Section \ref{sec:2} has the angular velocity
\begin{equation}
\begin{split}
\omega=\frac{a\Xi\left[\left(r^2+a^2\right)\Delta_\theta+\Delta_r\right]}{\Sigma^2}.
\end{split}
\end{equation}
the amount entering the boundary thermodynamic calculation as a chemical potential is the difference between the angular velocity at event horizon and infinity
\begin{equation}
\begin{split}
\Omega=\omega_H-\omega_\infty=\frac{a\left(l^2-r_H^2\right)}{l^2\left(r_H^2+a^2\right)}.
\end{split}
\end{equation}
This is exactly the angular velocity of the rotating Einstein universe at infinity~\cite{11}. Then we define the chemical potential 
\begin{equation}
\begin{split}
\mu=\frac{\Omega}{T_0}.
\end{split}
\end{equation}
And we use the chemical potential to express the rotational parameter as 
\begin{equation}
\label{eq:3.27}
\begin{split}
a=\frac{1}{\mu l}\left[\pi \left(l^2-r_H^2\right)+\sqrt{\pi^2\left(l^2-r_H^2\right)^2-\mu^2 l^2 r_H^2}\,\right].
\end{split}
\end{equation}
In order to calculate \eqref{eq:3.23}, we introduce $x=r_H/l$, and then \eqref{eq:3.23} becomes
\begin{equation}
\label{eq:3.28}
\begin{split}
S_n(\mu)=\frac{n}{n-1}\,\frac{1}{T_0}\int_{x_n}^{x_1}S_{therm}(x,\mu)\,\partial_x T(x,\mu)dx.
\end{split}
\end{equation}
where $x_n$ is the largest root of the equation $T(x,\mu)=T_0/n$. Combine \eqref{eq:3.27} with the entropy and temperature of the black hole we get 
\begin{equation}
\begin{split}
S(x,\mu)=\frac{V_H l^2}{4G_N}\frac{1-x^2+\Gamma_\mu(x)}{1-x^2+\Gamma_\mu(x)+\mu^2/{2\pi^2}}.
\end{split}
\end{equation}
\begin{equation}
\begin{split}
T(x,\mu)=\frac{T_0}{2x}\frac{\mu^2x^2/{\pi^2}-\left(x^2+1\right)\Gamma_\mu(x)+x^4-1}{x^2-1-\Gamma_\mu(x)}.
\end{split}
\end{equation}
Where the definition of the function $\Gamma_\mu(x)$ is
\begin{equation}
\begin{split}
\Gamma_\mu(x)\equiv\sqrt{1-\left(2+\mu^2/{\pi^2}\right)x^2+x^4}.
\end{split}
\end{equation}
To simplify the calculation, we integrate \eqref{eq:3.28} by parts
\begin{equation}
\begin{split}
&\int_{x_n}^{x_1}S_{therm}(x,\mu)\,\partial_x T(x,\mu)dx\\
&=\left.S_{therm}(x,\mu)T(x,\mu)\right|_{x_n}^{x_1}-\int_{x_n}^{x_1}T(x,\mu)\,\partial_x S_{therm}(x,\mu)dx.
\end{split}
\end{equation}
\begin{figure}[tbp]
\centering
\subfigure[ ]
{\includegraphics[width=.48\textwidth]{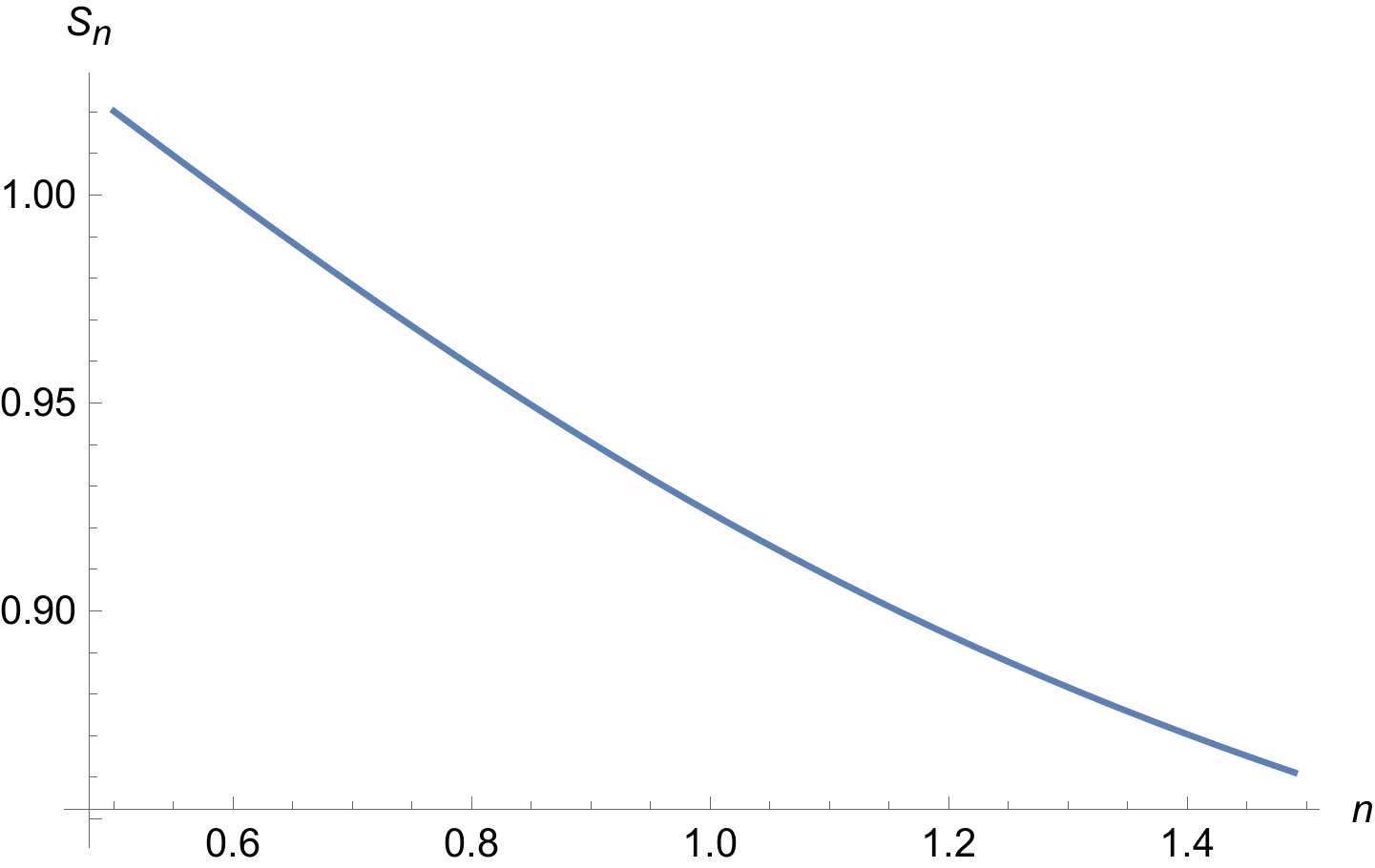}
\label{fig:1.a}
\hfill}
\subfigure[ ]
{\includegraphics[width=.48\textwidth]{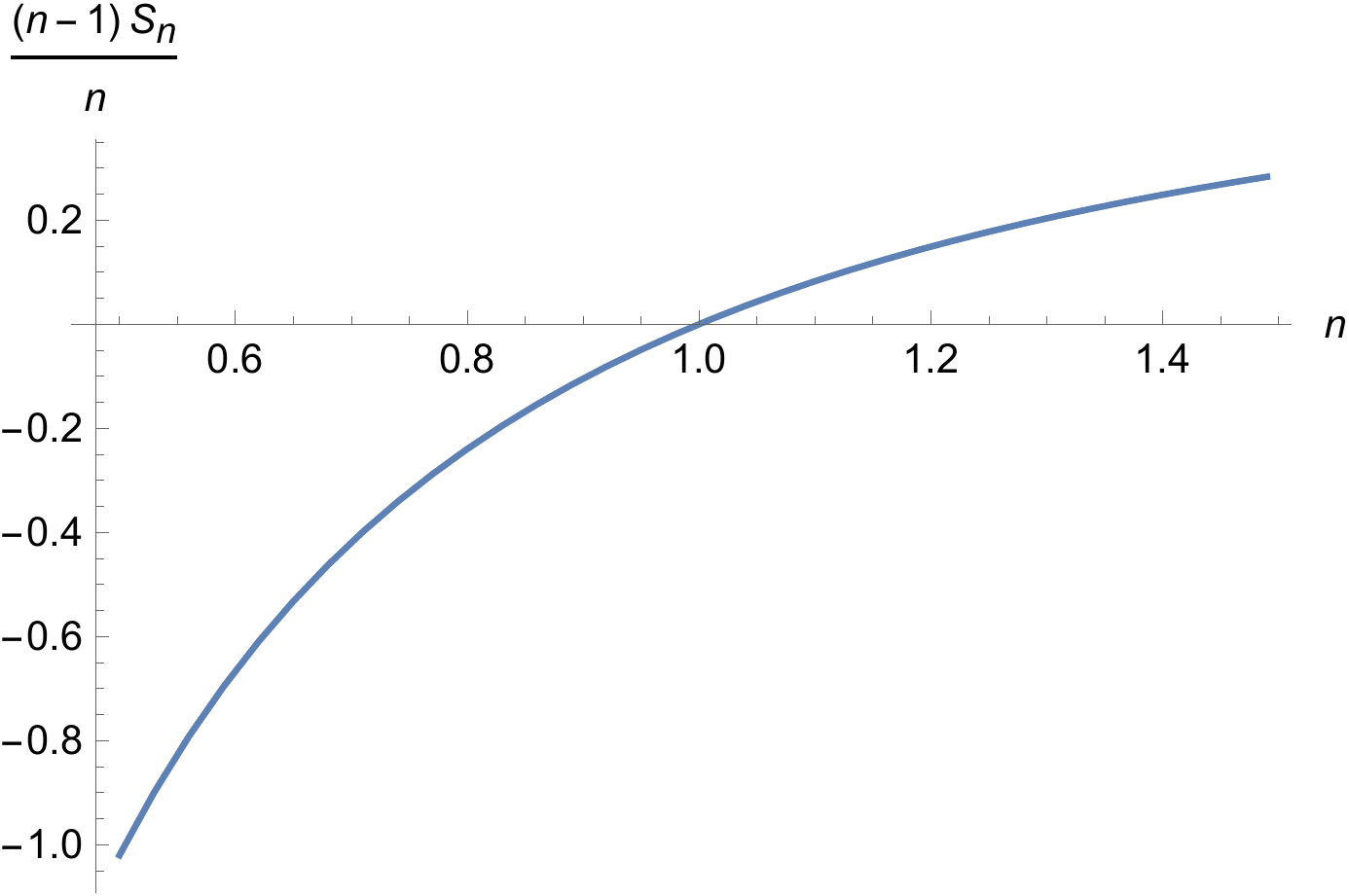}
\label{fig:1.b}}
\subfigure[ ]
{\includegraphics[width=.70\textwidth]{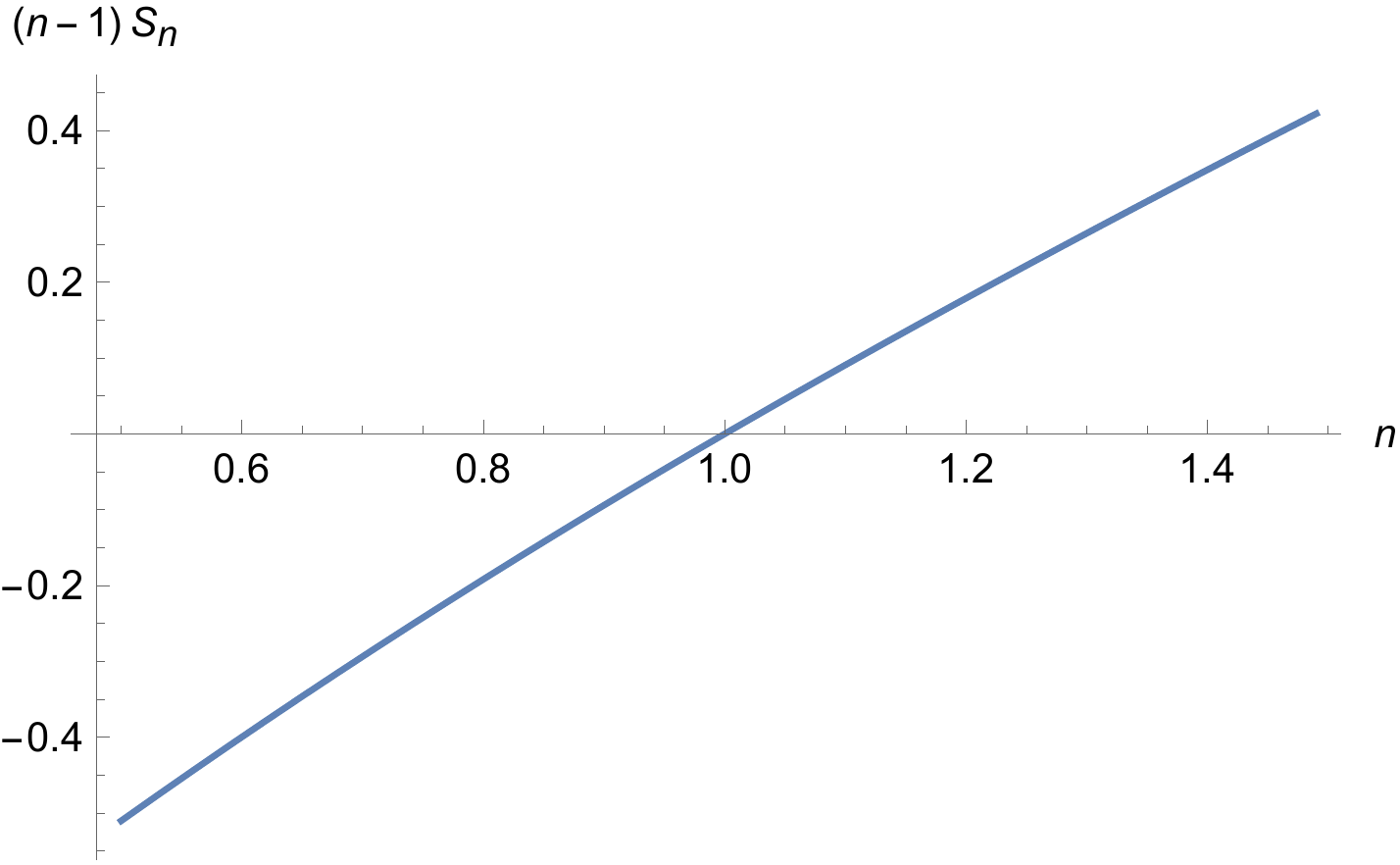}
\label{fig:1.c}}
\caption{\label{fig:1} Numerical results for the inequalities.}
\end{figure}
Combining these expressions then yields
\begin{equation}
\begin{split}
S_n(\mu)=\frac{V_H l^2}{8G_N}\frac{n}{n-1}\left\lbrace\frac{x_1^4-1-\left(x_1^2+1\right)\Gamma_\mu(x_1)}{x_1\left[x_1^2-1-\mu^2/{2\pi^2}-\Gamma_\mu(x_1)\right]}-\frac{x_n^4-1-\left(x_n^2+1\right)\Gamma_\mu(x_n)}{x_n\left[x_n^2-1-\mu^2/{2\pi^2}-\Gamma_\mu(x_n)\right]}\right\rbrace.
\end{split}
\end{equation}
We know that standard R\'{e}nyi entropies must satisfy a variety of different inequalities, in the following, we consider four such inequalities
\begin{equation}
\label{eq:3.34}
\begin{split}
\frac{\partial S_n(\mu)}{\partial n}\leq 0.
\end{split}
\end{equation}
\begin{equation}
\label{eq:3.35}
\begin{split}
\frac{\partial}{\partial n}\left[\frac{n-1}{n}S_n(\mu)\right]\geq 0.
\end{split}
\end{equation}
\begin{equation}
\label{eq:3.36}
\begin{split}
\frac{\partial}{\partial n}\left[(n-1)S_n(\mu)\right]\geq 0.
\end{split}
\end{equation}
\begin{equation}
\label{eq:3.37}
\begin{split}
\frac{\partial^2}{\partial n^2}\left[(n-1)S_n(\mu)\right]\leq 0.
\end{split}
\end{equation}
In particular, according to the entropies found in our holographic calculations, $i.e.$, the relation which we constructed between the R\'{e}nyi entropies for a spherical entangling surface and the thermal ensemble on the rotating hyperbolic background, these inequalities are naturally satisfied, as long as the thermal ensemble is stable, which the discussions could be found in ref.~\cite{12}. The results for the numerical calculations are given in the following figures.\\
\begin{figure}[tbp]
\centering
\includegraphics[width=.80\textwidth]{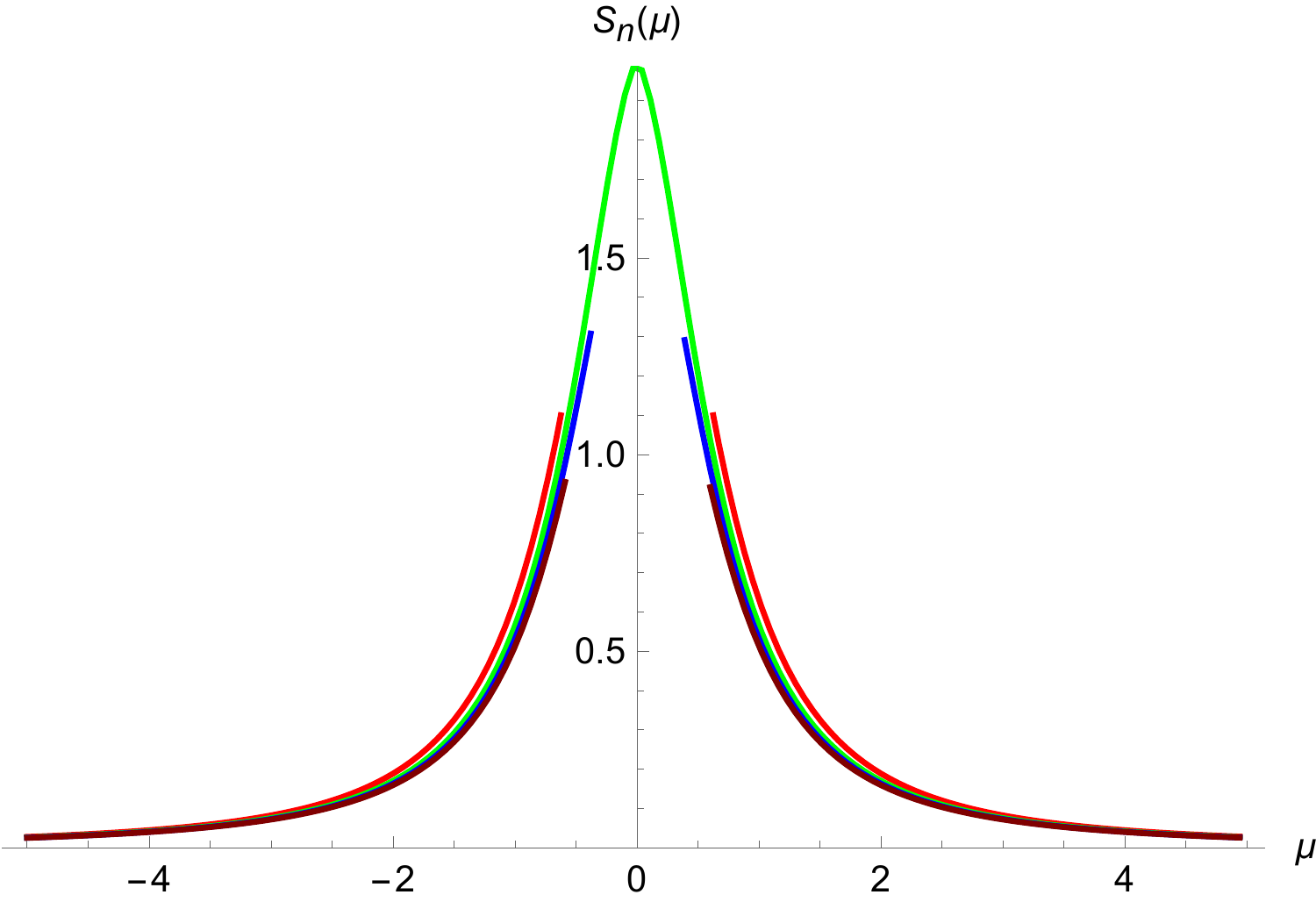}
\caption{\label{fig:2}Behaviors with chemical potential. The red curve corresponds to n=0.5, the green curve corresponds to n=1.0, the blue curve corresponds to n=1.5, and the brown one with n=2.0.}
\end{figure}
In Figure \ref{fig:1.a}, we plot $S_n(\mu)$ as a function of replica parameter $n$, we can clearly see that the first inequality satisfied. In Figure \ref{fig:1.b}, we plot $(n-1)S_n(\mu)/n$ as the function of $n$, it is obvious to see the second inequality is satisfied. In Figure \ref{fig:1.c}, we plot $(n-1)S_n(\mu)$ as the function of $n$, because the slope of the curve is positive so the third inequality is satisfied, and because the curve is convex so the fourth inequality also satisfied. And in Figure \ref{fig:2} we plot the rotating R\'{e}nyi entropy as the function of the chemical potential $\mu$ with difference choices of $n$, we found that when $|\mu|\to\infty$, the R\'{e}nyi entropy will turn to zero.\\
Note that \eqref{eq:3.23} can be expressed equally as
\begin{equation}
\begin{split}
S_n(\mu)=\frac{n}{n-1}\,\frac{1}{T_0}\left[F(T_0,\mu)-F(T_0/n,\mu)\right].
\end{split}
\end{equation}
Where $F(T,\mu)$ is the free energy in the boundary thermal CFT, one can easily find that
\begin{equation}
\begin{split}
F(T_0/n,\mu)=-T_0\frac{V_H l^2}{8G_N}\frac{x_n^4-1-\left(x_n^2+1\right)\Gamma_\mu(x_n)}{x_n\left[x_n^2-1-\mu^2/{2\pi^2}-\Gamma_\mu(x_n)\right]}.
\end{split}
\end{equation}
in the flat CFT, the free energy and the partition function have the relation
\begin{equation}
\begin{split}
F_n(\mu)=-\mathrm{log}\,Z_n(\mu).
\end{split}
\end{equation}
According to the relation between the thermal free energy and flat free energy
\begin{equation}
\begin{split}
F_n(\mu)=\frac{F(T_0/n,\mu)}{T_0/n}.
\end{split}
\end{equation}
From all of these, we can get the free energy and partition function on the n-fold cover of the space
\begin{equation}
\begin{split}
F_n(\mu)=-n\frac{V_H l^2}{8G_N}\frac{x_n^4-1-\left(x_n^2+1\right)\Gamma_\mu(x_n)}{x_n\left[x_n^2-1-\mu^2/{2\pi^2}-\Gamma_\mu(x_n)\right]}.
\end{split}
\end{equation}
\begin{equation}
\begin{split}
Z_n(\mu)=\mathrm{exp}\left\lbrace n\frac{V_H l^2}{8G_N}\frac{x_n^4-1-\left(x_n^2+1\right)\Gamma_\mu(x_n)}{x_n\left[x_n^2-1-\mu^2/{2\pi^2}-\Gamma_\mu(x_n)\right]}\right\rbrace.
\end{split}
\end{equation}
They are in the CFT with a chemical potential $\mu$ and temperature $\beta=2\pi n$.

\section{Conclusions}
\label{sec:4}
We have constructed a new class of entanglement measures which generalize the usual definition of R\'{e}nyi entropy with a chemical potential for a conserved angular momentum. For the special case that a CFT vacuum state with a spherical entangling surface, we can apply a conformal transformation to relate the rotating R\'{e}nyi entropies to the thermal entropies of a grand canonical ensemble, which is on the boundary of a certain rotating topological black hole discussed in Section \ref{sec:2}. \\
In section \ref{sec:3}, we considered the calculations of rotating R\'{e}nyi entropies using holography, where they are related to the thermal entropy of rotating topological black holes with hyperbolic horizons. We find that the rotational R\'{e}nyi entropy in the holographic model satisfies various inequalities, which were originally established for the standard R\'{e}nyi entropy without chemical potential. According to ref.~\cite{12}, we believe that the stability of the grand canonical ensemble in hyperbolic geometry is sufficient to ensure that the rotational R\'{e}nyi entropy satisfies these inequalities. All these results are also supported by numerical calculations.

\acknowledgments

I would like to thank Prof. Yang Zhou of the Department of physics of Fudan University for his guidance and the discussions with postdoctoral fellow Yu-sen An of the research group. They have provided very useful references and suggestions for the author.\newpage


\end{document}